\documentstyle[preprint,aps,epsf]{revtex}
\tighten

\begin{document}

\draft

\title{Modern nuclear force predictions for the neutron-deuteron 
scattering lengths
}

\author{
H.~Wita{\l}a$^{1,2}$,
A.~Nogga$^3$,
H.~Kamada$^4$,
W.~Gl\"ockle$^5$,
J.~Golak$^{1}$,
R.~Skibi\'nski$^1$.
}
\address{$^1$M. Smoluchowski Institute of Physics, Jagiellonian University,
                    PL-30059 Krak\'ow, Poland}
\address{$^2$ Department f\"ur Physik und Astronomie, Universit\"at Basel, 
 Basel, Switzerland}
\address{$^3$ Department of Physics, University of Arizona, Tucson,
              Arizona 85721}
\address{$^4$ Department of Physics, Faculty of Engineering,
   Kyushu Institute of Technology,
   1-1 Sensuicho, Tobata, Kitakyushu 804-8550, Japan}
\address{$^5$Institut f\"ur Theoretische Physik II,
         Ruhr-Universit\"at Bochum, D-44780 Bochum, Germany}

\date{\today}
\maketitle

\begin{abstract}
The neutron-deuteron (nd) 
doublet ($^2a_{nd}$) and quartet ($^4a_{nd}$) scattering lengths 
have been
calculated based on the nucleon-nucleon (NN) 
interactions CD~Bonn~2000, AV18, Nijm~I, II and 93 
alone and in selected
combinations with the Tucson-Melbourne (TM), a modified version thereof, TM99, 
and the Urbana IX three-nucleon (3N) forces. 
For each NN and 3N force combination also 
the $^3H$ binding energy was calculated. 
In case of TM99 and Urbana IX 
the 3NF parameters were adjusted to the $^3$H binding
energy. In no case (using np-nn forces) the experimental value of $^2a_{nd}$
was reached. We
also studied the effect of the electromagnetic interactions in the form
introduced in AV18.
Switching them off for the various nuclear force models leads to shifts of up to
+0.04 fm for $^2a_{nd}$, which is significant for present day standards. 
The electromagnetic effects have
also a noticeable effect on $^4a_{nd}$, which otherwise is extremely stable under the
exchange
of the nuclear forces. Only when the electromagnetic interactions are included
the current
nuclear forces describe the experimental value. As a consequence of the failure to
reproduce $^2a_{nd}$ also the newly measured coherent nd scattering length 
($b_{nd}$) can not be reproduced. For the current nuclear force models 
there is a strong scatter of the $^3$H
binding energy and the  $^2a_{nd}$ values around an averaged straight line 
(Phillips line). This allows to use $^2a_{nd}$ and the $^3$H binding
energy as independent low energy observables.
\end{abstract}

\narrowtext

\section{Introduction}
\label{secII}

It has been observed a  long time  ago that the nd scattering length for total
three-nucleon spin S=1/2 ($^2a_{nd}$) is correlated to 
the $^3H$ binding energy ($E_{^3H}$). This
correlation  is known as
the Phillips line~\cite{phil68}. Indeed, calculations years 
later based on simplistic or more realistic
 NN  model forces (see~\cite{fri83,fri84,fri86,chen89,kiev94,kiev97}) yielded quite a 
few results for the $^3H$ binding energy and the $^2a_{nd}$
scattering length, which lie on or 
rather close to a line in the  two-dimensional
plane spanned by $E_{^3H}$ and
$^2a_{nd}$. Also 3N forces of the $2\pi$-exchange type 
have been added. In~\cite{fri84} it was found  
that this line passes well through the
experimental point.

In recent years chiral perturbation theory and effective 
theories have been applied to
nuclear physics. In the pionless formulation~\cite{beda98,beda02,beda02_a}, 
adequate for extreme low 
energy phenomena, it
has been shown 
that  $^3H$ can be energetically stabilized only if a 
3N contact force
 is introduced (see however~\cite{xxx,xxx1}).
In the two lowest orders of that framework there is just 
one parameter connected to that 3N force. Thus both quantities, 
$E_{^3H}$ and $^2a_{nd}$, depend on that one parameter 
and are therefore
correlated though the line does not hit the experimental point. 
In higher orders additional parameters show up and the 
correlation is broken, which makes the two quantities independent. The same 
observation was made
in an approach  based on 
chiral perturbation theory~\cite{epel02} which includes 
explicitely the pion
degrees of freedom. In the  next-to-next-to-leading-order  
(NNLO), 3N forces occur the 
first time and they  depend on two
parameters. This makes $E_{^3H}$ and $^2a_{nd}$ independent and 
the Phillips line correlation is
broken. In fact in that framework  the two experimental values 
are used to fix those two
parameters of the 3N force~\cite{epel02}. Thus we find it interesting to ask,  
whether  the
 conventional and
high precision NN forces AV18~\cite{av18}, CD~Bonn~2000~\cite{cdb2000}, 
Nijm~I, Nijm~II, and Nijm~93~\cite{nijm} 
alone or in combination with the 
two  most popular  3N force models , Urbana~IX~\cite{uix}
 and TM99~\cite{fri99,con01} 
 (an updated Tucson-Melbourne $2\pi$-exchange
3NF~\cite{tm} modified in view  of 
chiral symmetry) lead to a strict 
correlation between $E_{^3H}$ and
$^2a_{nd}$ or whether that Phillips line correlation is also broken. 
  Further we ask whether the NN and 3NF combinations adjusted 
to $E_{^3H}$ (or may be only one of them)
also describe $^2a_{nd}$. One more reason to confront 
$^2a_{nd}$ to state-of-the-art calculations is the recent
appearance of a precision neutron 
interferometric measurements of the nd coherent
scattering length ($b_{nd}$)~\cite{black02}. 

The coherent scattering length $b_{nd}$ depends in 
addition to $^2a_{nd}$ also on the second s-wave scattering 
length for the state of total 3N spin S=3/2 ,
$^4a_{nd}$ . Because of the Pauli principle this quantity is supposed not 
to be sensitive to  
short range details of nuclear forces. We want to investigate 
also that quantity
in the
light of modern  nuclear forces.

Then we would like to add two more investigations. 
Charge-symmetry breaking (CSB) in the strong NN forces is mostly
pronounced in the states $^1S_0$, where the scattering length for the  
neutron-neutron (nn), $a_{nn}$, and proton-proton (pp), $a_{pp}$, systems are
different. However the value for $a_{nn}$ is still under 
debate~\cite{trot99,huh00}.
Therefore we would like to
present results, where the nn forces are replaced by the (strong) pp 
forces. This will
provide some insight into the magnitudes of the  shifts in $^2a_{nd}$ caused 
by small changes in $a_{nn}$.

The other investigation is due to effects on the scattering lengths 
and $E_{3H}$ caused by electromagnetic interactions, mostly due to 
magnetic moment interactions (MMI). MMI is a relativistic effect 
and including only that
specific force is of course inconsistent, since other relativistic 
effects are not taken into
account (see for instance~\cite{kam02} and references therein). But it is 
interesting to see its
separate effect on $^2a_{nd}$ and 
$^4a_{nd}$ (its effect on the binding energy of $^3H$ and $^3H$e is known and
older results have been reconfirmed recently~\cite{nogga03}). 
 Here we hit some "defects" 
in current NN force
models. The NN potentials CD~Bonn~2000, Nijm~I, II, and Nijm~93 are fitted 
directly to the NN data without taking 
electromagnetic interactions (EMI) into account 
(of course the point Coulomb 
force in case of the
pp system has been included). Therefore those strong forces include
the effects of the
MMI's (and further electromagnetic corrections). If we want to see then 
the separate
effects of the EMI's we have to subtract from those model NN forces 
the EMI's and compare to results without that subtraction. In
case of AV18 the strong force plus EMI's have 
been fitted to the data. Thus the force free of EMI's is just the 
strong AV18 force alone. In this manner by comparing results with and without 
electromagnetic interactions 
one can see
their effects.

The paper is organised as follows. In Section~\ref{secIII}  
the theoretical formulation is  briefly outlined. 
The results are given in Section~\ref{secIV}, 
and we end with a summary and an outlook in Section~\ref{secV}. More 
technical details are deferred to an appendix. 

\section{Formulation}
\label{secIII}

We use the Faddeev scheme. Including a three-nucleon force 
a convenient basic formulation for
part of the $nd \to n+n+p$ break-up amplitude $T$~\cite{Report,hub97} 
is the integral equation 
\begin{eqnarray}
T &=& tP\phi + (1+tG_0)V_4^{(1)}(1+P)\phi + tPG_0T + 
(1+tG_0)V_4^{(1)}(1+P)G_0T.
\label{eq:1}
\end{eqnarray}

The driving term contains the NN $t$-operator $t$, permutation operators $P$, 
the free 3N
propagator $G_0$ and part of the 3N force,  $V_4^{(1)}$. Any 3N force can 
be split into three pieces, 
where for instance the first piece  is symmetrical under exchange 
of particles 2 and 3 , the
second under 3-1 exchange etc. Thus the quantity $V_4^{(1)}$ is 
the part symmetrical under 2-3 exchange
like the $t$-operator  $t$, which  is supposed to act on  the pair 23. 
Finally, $\phi$ is
 the initial channel state
composed of the deuteron state and a momentum eigenstate of the 
projectile neutron.That
integral equation can be solved precisely in a partial wave 
decomposition in momentum
space. For details see~\cite{glo83,Report,corn}. 

The operator $U$ for elastic scattering 
is given in terms of
the amplitude $T$ by quadrature as follows
\begin{eqnarray}
U &=& PG_0^{-1}\phi + PT + V_4^{(1)}(1+P)\phi + V_4^{(1)}(1+P)G_0T.
\label{eq:2}
\end{eqnarray}

We want to solve the integral equation (1) directly at  the 
threshold of nd scattering.
 This
is for zero initial relative momentum $\overrightarrow q_0$  of 
the projectile and 
will then lead directly to
the scattering length. For the convenience of the reader we briefly 
sketch the necessary
steps~\cite{huber}. Our partial wave momentum space basis is 
denoted by $\vert pq\alpha >$, where p and q are the
magnitudes of standard Jacobi momenta and $\alpha$ a string of 
angular momentum 
and isospin quantum  numbers (see~\cite{Report,glo83}). 
Then for $\overrightarrow q_0$ 
  in z- direction 
we define the auxiliary
amplitude
\begin{eqnarray}
U_{\alpha,{\lambda}I}(p,q) &=& \sum_{m,m_d} 
{{\sqrt{4\pi\hat{\lambda}}}\over{\hat{J}} }(\lambda~0~{{1}\over{2}}m\vert Im)
(j_dm_dIm\vert Jm_d+m)<pq\alpha \vert U \vert \phi >
\label{eq:3}
\end{eqnarray}
for the projectile nucleon with orbital angular momentum 
$\lambda$ ($\hat{\lambda} \equiv 2\lambda + 1$) and total angular momentum $I$
combined with the deuteron total angular momentum $j_d =1$ to total 3N angular 
momentum $J$. 
Out of that amplitude one obtains the partial wave projected 
nd elastic scattering
amplitude
\begin{eqnarray}
U_{{\lambda}'I',{\lambda}I}^{J} &=& 
\sum_{l'} \int {p'}^2dp' \phi_{l'}(p') 
U_{{{\alpha}'}_d,{\lambda}I}(p',q_0).
\label{eq:4}
\end{eqnarray}
where $\phi_{l}(p)$ are the s- and d-wave components of the deuteron and 
${\alpha}_d$ contains the deuteron quantum numbers. 

Finally, the projectile spin and the deuteron spin can be 
combined to the total spin $\Sigma$ 
and one obtains
\begin{eqnarray}
U_{{\lambda}'{\Sigma}',{\lambda}{\Sigma}}^{J} &=& 
\sum_{I,I'} \sqrt{{\hat{\Sigma}'}{\hat I}'} (-)^{J-I'} 
\biggl{\lbrace} {{\lambda}'\atop {j_d}} {{1\over 2}\atop J} 
{{I'}\atop {\Sigma}'} \biggr{\rbrace}
\sqrt{{\hat{\Sigma}}{\hat I}} (-)^{J-I} 
\biggl{\lbrace} {{\lambda}\atop {j_d}} {{1\over 2}\atop J} 
{{I}\atop {\Sigma}} \biggr{\rbrace}
U_{{\lambda}'I',{\lambda}I}^{J}.
\label{eq:5}
\end{eqnarray}

The $S$- matrix element is then given as
\begin{eqnarray}
S_{{\lambda}'{\Sigma}',{\lambda}{\Sigma}}^{J} &=& 
{\delta}_{{\lambda}'{\lambda}}{\delta}_{{\Sigma}'{\Sigma}} 
-i{{4\pi}\over{3}}mq_0(i)^{{\lambda}'-\lambda}
U_{{\lambda}'{\Sigma}',{\lambda}{\Sigma}}^{J}
\label{eq:6}
\end{eqnarray}
leading to the doublet and quartet  scattering lengths for $q_0=0$
\begin{eqnarray}
^2a_{nd} &=& {{2\pi}\over{3}}~ m~ 
U_{0{1\over 2},0{1\over 2}}^{{1\over 2}}  \cr
^4a_{nd} &=& {{2\pi}\over{3}}~ m~ 
U_{0{3\over 2},0{3\over 2}}^{{3\over 2}}.
\label{eq:7}
\end{eqnarray}

One also defines a coherent scattering length $b_{nd}$ as
\begin{eqnarray}
b_{nd} &=& {{m_n+m_d}\over{m_d}}~ \lbrack~ ({1\over 3})~^2a_{nd} + 
({2\over 3})~ ^4a_{nd} ~\rbrack
\label{eq:8}
\end{eqnarray}

We defer the special form of the Faddeev integral equation (1) 
at $q_0=0$ to the appendix.
It is free of singularities and therefore as easily solved as a 
bound state problem. Also
the explicit form of the elastic amplitude              for $q_0=0$ is 
given there.

\section{Results}
\label{secIV}

We used the NN forces CD~Bonn~2000, AV18, Nijm~I, II and Nijm~93 alone 
or various combinations
with the three-nucleon forces Urbana~IX, 
the older Tucson-Melbourne force  and
 the modified one TM99. 
When we combine the 
Urbana IX 3NF with CD~Bonn~2000 the strength of the repulsive 
part of this 3NF has been reduced by multiplying it with the factor 0.812 
in order to get the proper $E_{^3H}$. 

Due to their nonnegligible influence
on the nd scattering lengths, we took special care of electromagnetic
interactions. 
In the case of the AV18 potential it is 
clear how to separate the strong AV18 force from the 
electromagnetic parts because both are well defined and added together in fitting
the total force to the NN data. 
In case of the np system the EMI's are given in Eqs. (11), (12), and (15) of
Ref.~\cite{av18} and
in Eq. (16) for the nn system (for the np system we 
did not  include the very small class IV charge asymmetric 
force $\propto \overrightarrow L \cdot 1/2(\overrightarrow {\sigma}_i - 
\overrightarrow{\sigma}_j)$. Also we neglected 
the energy dependence of the ${\alpha}'$).  
This is different for
the CD~Bonn~2000 and Nijmegen potentials, which  
were fitted directly to the NN data without adding to them electromagnetic 
interactions, with the exception of the point Coulomb force in case of the 
pp system. Therefore to define the strong forces  in the particular NN system 
one needs to subtract the corresponding EMI, which we assumed to be given 
as in ref.~\cite{av18}. 
To be precise for the np system we subtract from the np CD Bonn and the 
Nijmegen
forces the np EMI's as defined above. Similarly for the nn system we subtract
from CD
Bonn the MMI as defined above. Since we also want to see the effect 
of replacing
the
strong nn force by the strong pp force we have also to define the strong pp CD
Bonn and
Nijmegen forces. In this case we subtract from those potentials  the 
pp EMI's as
given in
Eqs. (3)-(8) of Ref.~\cite{av18} without the leading 1 in $F_c (r)$ from 
Eq.~(10) 
of Ref.~\cite{av18},
since the
point Coulomb force has been taken into account for those potentials fitting to
the pp
data.

\begin{figure}[h!]
\leftline{\mbox{\epsfysize=180mm \epsffile{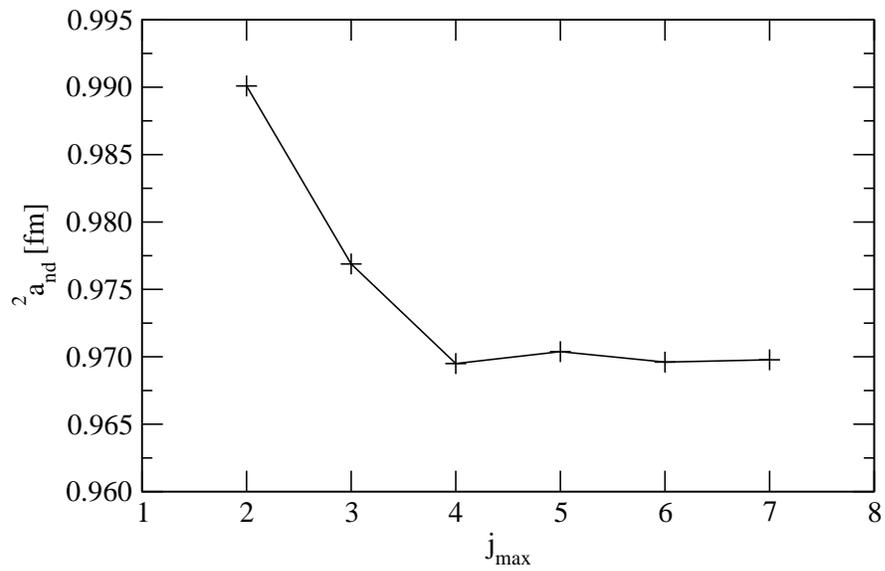}}}
\caption[ ]
{
The convergence of  the doublet scattering length $^2a_{nd}$ as a 
function of the 2N total angular momentum $j_{max}$ for the CD~Bonn potential.
}
\label{fig1}
\end{figure}

\begin{figure}[h!]
\leftline{\mbox{\epsfysize=180mm \epsffile{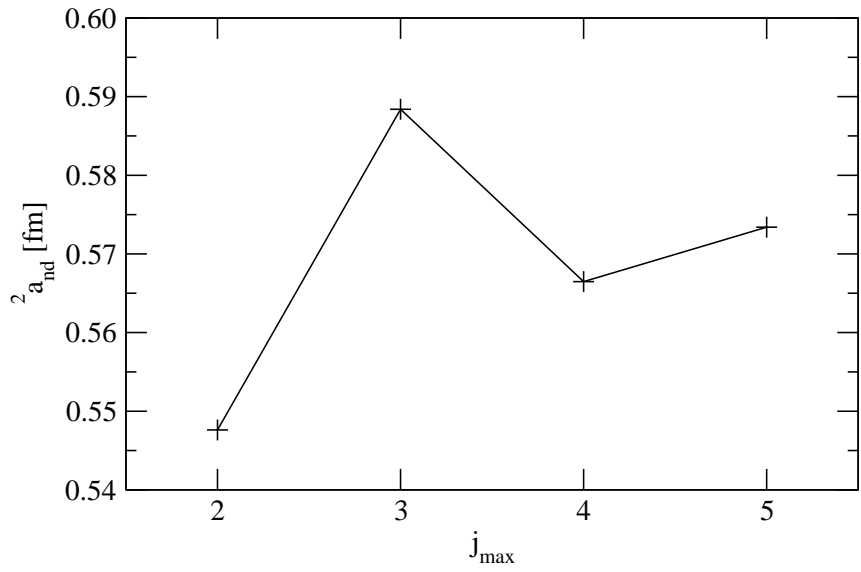}}}
\caption[ ]
{
The same as in Fig.~\ref{fig1} for the CD~Bonn potential 
combined with the TM 3NF.
}
\label{fig2}
\end{figure}

\begin{figure}[h!]
\leftline{\mbox{\epsfysize=180mm \epsffile{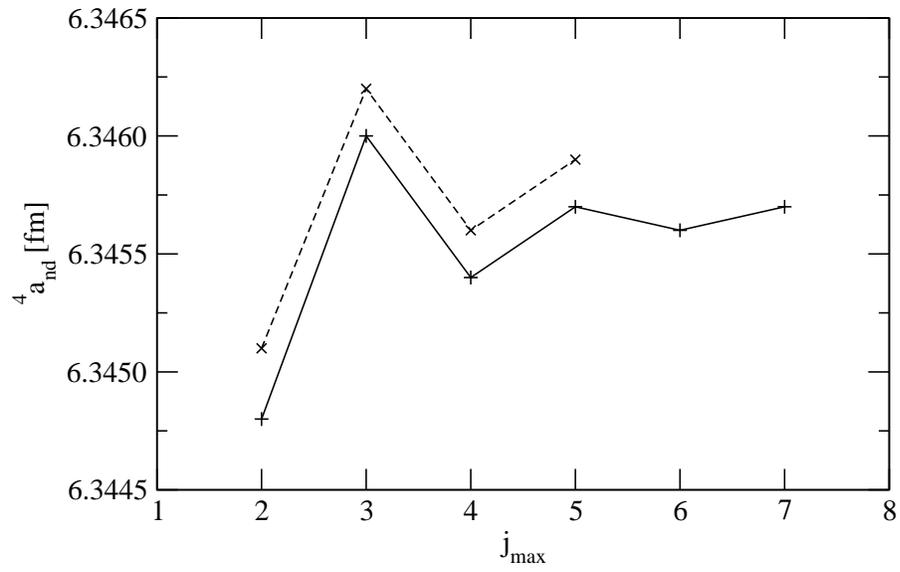}}}
\caption[ ]
{
The convergence of  the quartet scattering length $^4a_{nd}$ as a 
function of $j_{max}$ for the CD~Bonn 
potential (solid curve) 
and its combination  with the TM 3NF (dashed curve).
}
\label{fig3}
\end{figure}

Before we report on our results we give some comments on our 
numerical accuracy. As usual the
partial wave decomposition is truncated at a certain total 
two-body angular momentum $j_{max}$. Fig.~\ref{fig1} documents 
the convergence of $^2a_{nd}$ as function of $j_{max}$. 
This shows that we reached
an accuracy of about three digits. This is for CD~Bonn~\cite{cdb} alone and 
valid also for the other
NN forces alone. Adding a three-nucleon force we were limited 
to $j_{max}=5$ due to
computer resources. Nevertheless, as Fig.~\ref{fig2} documents, the convergence 
reached for $^2a_{nd}$ is
 two digits. In case of $^4a_{nd}$ with NN forces alone 
we reach 4 digits convergence
and including a 3N force  an accuracy  close to that. This 
is documented in Fig.~\ref{fig3}. The
 other
numerical ingredients (discretization in the momenta) are 
safely under control. In all calculations we took into account charge 
dependence of the NN forces using a simple ``${2\over 3}t_{pp(nn)} + 
{1\over 3}t_{np}$'' rule to generate t-matrices in isospin $t=1$ 
2N states~\cite{cdbr}. The total isospin $T=3/2$ 3N states 
have been neglected~\cite{cdbr}. We checked that their inclusion does not 
change $^4a_{nd}$ up to the fifth  digit and the 
change of $^2a_{nd}$ is of the 
order of $0.1\%$. The triton binding energies have been obtained using 
$j_{max}=6$. They are accurate to $2$~keV.  

\begin{figure}[h!]
\leftline{\mbox{\epsfysize=180mm \epsffile{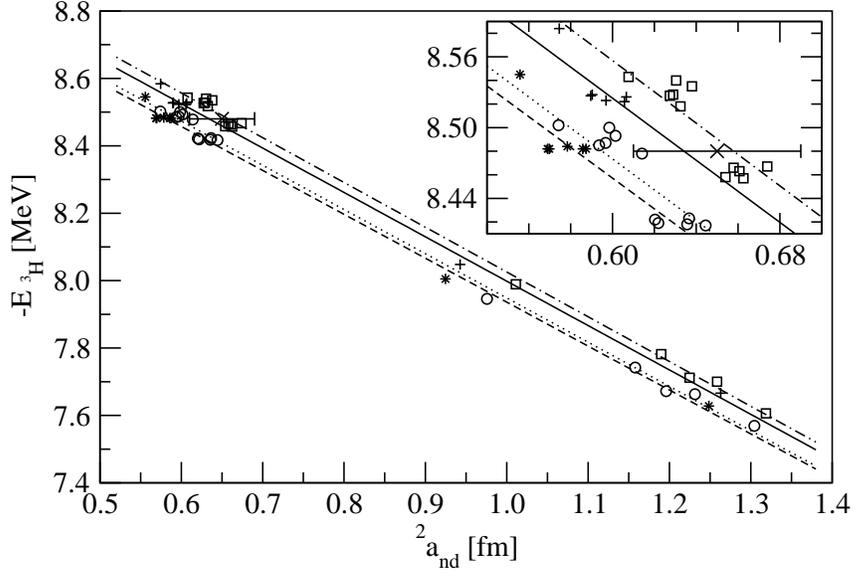}}}
\caption[ ]
{The results for $^2a_{nd}$ and $E_{^3H}$ from Table~\ref{tab1}: 
np-nn forces alone (pluses), np-pp forces alone (squares),  and  
np-nn and np-pp forces 
plus electromagnetic 
interactions (stars and circles, respectively). The four straight lines 
(Phillips lines) are $\chi^2$-fits 
(np-nn: solid, np-pp: dashed-dotted, np-nn with EMI's: dashed, 
np-pp with EMI's: dotted). The lines with EMI's 
 miss the experimental error 
bar for $^2a_{nd}$~\cite{exp1}. The physically interesting domain around the 
experimental values is shown in the inset.
}
\label{fig4}
\end{figure}

As an overview we show all our results for $^2a_{nd}$ and $E_{^3H}$ in 
Fig.~\ref{fig4}.  
We see a group of
results in the right half of the figure based on NN forces alone 
and another group close to
the experimental area including 3N forces. 
 We performed several investigations. First 
we take CD~Bonn~2000 as it is (fitted to the NN data) and use the 
np-nn force
combinations appropriate to the nd system. 
In this way the EMI's in the np and nn systems are effectively included
inside the strong forces. 
In case of AV18 
 we keep all electromagnetic corrections as in~\cite{av18} 
except the energy dependence of ${\alpha}'$ (MMI's for 
the nn system and the MMI's 
together with the one photon Coulomb term $V_{C_1}(np)$ for 
the np system). The corresponding two predictions are shown as stars in 
Fig.~\ref{fig4}.

Secondly we 
want to see the effect of replacing the strong nn forces by the 
strong pp forces.
The difference between nn and pp strong forces is mostly located in
the different scattering lengths $a_{nn}$ and $a_{pp}$ (strong) and  will
therefore give some
information how changes in $a_{nn}$
will show up in changes  of $^2a_{nd}$.
Since thereby we do not want to change the EMI's we keep in case of AV18 the nn
MMI. For the CD Bonn and the  Nijmegen potentials the strong pp potentials are
defined as above and
the nn MMI (as for AV18) is added. The results are shown as the 5 open circles
in Fig 4.
As seen in case of CD Bonn 2000 and AV18 the $^3$H binding energy is 
decreased and
$a_{nd}$ increased. For the Nijmegen potentials such a comparison is not done, 
since no nn forces have been introduced.
The effects on the 3N
binding energies are  known. 
These two first investigations provide theoretical predictions for the 
nd scattering lengths and triton binding energy including all electromagnetic 
interactions similarly as is the case for 
the measured values. 

In the next two investigations 
we address the effects of the electromagnetic interactions 
themselves by switching 
them off while generating theoretical predictions. For the AV18 potential 
we take just 
the np-nn and np-pp strong force combinations alone, while in the cases 
of CD~Bonn~2000, Nijm~I, II, and 93 we only use the   
corresponding strong forces 
stripped off from the EMI's, as described  
above. The resulting 
theoretical predictions are shown as pluses and squares  in Fig.~\ref{fig4} for 
the np-nn and np-pp combinations, respectively. 
Again the binding energy is decreased and $a_{nd}$ increased.

The individual results of these four investigations are
summarized  in Fig.~\ref{fig4} also as dashed (np-nn with EMI's), 
dotted (np-pp with EMI's), solid (np-nn), and dashed-dotted (np-pp)  
straight lines fitted in a $\chi^2$
sense.
These lines include also the corresponding results including 3NF's (see below).
We see a small shift of the lines  under exchanges of nn versus pp forces, 
but a more significant shift
if  the electromagnetic  forces are switched off. 
Though the two curves (dashed and dotted) for the cases when the 
electromagnetic 
forces are added come close to the experimental range spanned by the 
uncertainty in $^2a_{nd}$, they miss it clearly. When the electromagnetic 
forces are switched off the np-nn (solid) and np-pp (dashed-dotted) 
lines go through the experimental point well inside the $^2a_{nd}$ 
error bar. 

Now we want to regard our results in more detail as displayed 
in Table~\ref{tab1} and in the
inset of Fig.~\ref{fig4}. The theory has to be finally compared  to the 
experimental values, which
are $^2a_{nd}=(0.65~\pm~0.04)$~fm~\cite{exp1}, 
$^4a_{nd}=(6.35~\pm~0.02)$~fm~\cite{exp1}, and 
$b_{nd}=(6.669~\pm~0.003)$~fm~\cite{black02}.

The results in Table~\ref{tab1} are grouped into NN force predictions 
only and selected combinations
with the 3N forces TM, TM99 and Urbana~IX. For each potential or 
potential combination we
show the  results for the various scattering lengths 
and the $^3$H binding energies. This is given for the np-pp NN forces, 
with (without)  EMI's in the first (second) row. 
For AV18 and CD~Bonn~2000 we also show  the results for np-nn forces with 
(third row) and without (fourth row) electromagnetic interactions. 
Note that in case of the np-nn forces including EMI's (as described above) the
combinations with TM99 and Urbana~IX are fitted well to the experimental 
value -8.48
MeV of
the $^3$H binding energy. This is also the case for Nijm~I and II, 
which, however,
refers to the np-pp forces.
For the older TM 3N force we did  not 
perform a precise  (re)fit
and the results are only included in view of  investigating,  
whether a straight line
correlation between $^2a_{nd}$ and $E_{^3H}$ exists. 
A glance at Fig.~\ref{fig4} 
tells that the individual
results scatter around the four straight lines. Thus obviously 
no straight line
correlation exist (this has been known before, though for some 
older calculations the numerical accuracy might be not sufficient 
to give a reliable judgement).

Let us now concentrate  on the group of results with 3N forces. 
These are displayed in the
inset of Fig.~\ref{fig4}. We see  four results (stars) for the np-nn forces 
including TM99 
or Urbana~IX, where the
binding energy has been exactly fitted but where the $^2a_{nd}$ value 
is too small. 
These are the results achieved under the supposedly  most realistic assumptions
in this paper.
If one switches off 
the electromagnetic interaction (pluses) the binding energy increases and 
interestingly $^2a_{nd}$ 
moves  to larger
values. 
Regarding all results, 
the inclusion of
the electromagnetic force in our studies  shows
that they cause
shifts of up to about 40 keV less binding energy and of up to about 0.04 fm
decrease in $^2a_{nd}$. In no case studied the experimental value of 
$^2a_{nd}$  is reproduced for np-nn and np-pp strong forces combined with 
different 3NF's with exception of np-pp AV18 combined with TM 3NF, for which 
the theoretical prediction lies at the lower limit of the error bar.

As one learned from the approach in chiral perturbation theory~\cite{epel02}, 
where two parameters are
needed to fix the short range 3N forces at NNLO and consequently two 3N
observables to adjust them, one could foresee that the straight 
lines in Fig.~\ref{fig4} could
only by accident pass through the experimental region. For the conventional 
forces used in this paper,  
one can think of additional 3N force diagrams (the most obvious 
one the $\pi-\rho$ exchange)
where a sufficient number of parameters would be available to fit both, 
$E_{^3H}$ and $^2a_{nd}$.

Going back to Table~\ref{tab1} we see that $^4a_{nd}$ sticks always 
close to the 
value  6.34  for the np-pp
and np-nn NN force choices, without or with 3N forces and with 
EMI's included. This is well within the 
experimental $^4a_{nd}$ error bar.  
Interestingly, the electromagnetic interactions 
increase $^4a_{nd}$ in nearly all 
cases by about 0.02 and the pure strong force  predictions 
lie always outside the experimental error bar.

Finally, one can confront
theory to the very precisely known experimental value of the 
coherent scattering length
$b_{nd}$~\cite{black02}.
Clearly the supposedly most realistic  dynamics (nn-np NN forces plus TM99 or
Urbana IX 3NF's) misses that value.
As can be seen from Table~\ref{tab1}, when 
electromagnetic interactions are included 
the  np-pp force combination reaches the experimental 
value  in case of the AV18 and CD~Bonn~2000 potentials combined 
with Urbana IX and AV18 with TM99. However, this agreement is accidental 
and caused by the corresponding decrease in $^3H$ binding.

\section{Summary and Outlook}
\label{secV}

A recently performed precise neutron interferometric measurement 
of the nd coherent
neutron scattering length~\cite{black02} and a planned 
precision measurement of the doublet nd
scattering length~\cite{plan} stimulated us to investigate the 
theoretical predictions of that quantity for the high
precision NN forces CD~Bonn~2000, AV18, Nijm~I, II and 93 in 
combination with currently  popular 3N
force models. These are the modified $2\pi$-exchange 
Tucson-Melbourne (TM99) and the Urbana~IX
3N forces. We have chosen several NN and 3NF combinations, which are separately
adjusted to the $^3H$ binding energy. For NN forces alone with 
and without EMI's we 
recovered the approximate correlation
between $E_{^3H}$ and $^2a_{nd}$, but the scatter 
around  a thought straight line (Phillips line) inside the band spanned 
by the 4
lines in Fig.~\ref{fig4} is quite strong. 
Adding 3N forces shifts 
the values into the neighbourhood of the
experimental range of $^2a_{nd}$, but misses 
the experimental value including its error bar
in all cases, when electromagnetic forces are included. 
The inset of Fig.~\ref{fig4} clearly shows that for 
equal or nearly equal $^3$H binding
energies $^2a_{nd}$ can vary significantly and vice versa.

Thus one has to conclude that $^2a_{nd}$ has to 
be considered as a low energy
observable, which is independent from the $^3H$ binding energy. 
This observation has been
found before in approaches based on pure  
effective field theory (pionless formulation) and
on chiral  perturbation theory (including pion degrees of freedom). 
Thus in future investigations, adjusting 
both observables, $E_{^3H}$ and $^2a_{nd}$, for conventional nuclear forces  
 will require  more flexibility in
the choice of 3N forces. Adding more mechanisms (on top of the 
$2\pi$-exchange) for 3N
forces  should be no obstacle. This is a step already 
performed in the effective
theory approaches~\cite{beda98,beda02,beda02_a,epel02}.

We also investigated the effects on $^2a_{nd}$ resulting from 
electromagnetic  interactions given
in~\cite{av18}. The  effects on $^2a_{nd}$ and even $^4a_{nd}$ 
are noticeable. 
For $^2a_{nd}$ including the electromagnetic interactions reduces its
value by up to 0.04 fm. 
It is interesting to note that $^4a_{nd}$ is 
perfectly stable under all
exchanges of nuclear forces studied in this paper but the 
electromagnetic interactions
affect its value, though only in the 3'rd digit. However, only when 
EMI's are included the experimental value is reproduced. 

The effects of adding the electromagnetic  interactions  on the $^3H$ 
binding energy are well
 known and can reach shifts of up to 40 keV 
less binding energy.

Due to the  failure to describe $^2a_{nd}$ also the
recently newly measured coherent scattering length $b_{nd}$ 
cannot  be reproduced
theoretically. The good 
reproduction of $^4a_{nd}$ by all interactions and 
the small error bar of 
the coherent scattering 
length suggests that the value of the doublet nd scattering length might 
be 
somewhat smaller than the presently one, namely around 0.63 fm,
different from 
the present experimental value of $^2a_{nd}$. This  
strongly calls for a new, more precise measurement. 

Since the scattering lengths are (extreme) low energy observables, it 
appears that the
mentioned effective theory approaches  are the  most adequate ones. 
Because one works there
below a certain momentum cut off, which is smaller than the nucleon mass, 
they allow also to
incorporate relativistic effects in a well defined and converged manner. 
Also 3N forces
appear in those approaches  in a well organised manner, according to 
a certain power
 counting scheme, and
are consistent to the NN forces. In other words, one can take 
into account all these subtle
effects, relativity, 3N forces, isospin breaking, in a well controlled 
and systematic
manner.  In conventional approaches on the other hand, 
which include a lot of
phenomenological  parametrisations and where no momentum cut-off 
is used, a reliable treatment
of relativistic effects poses still a problem and the choices of 
3N force mechanisms are
also
quite open. Therefore in conventional approaches 
physically reliable predictions  to $^2a_{nd}$ will very
likely remain a challenge for quite some time.  

\section{Appendix}

This appendix summarizes 
various expressions exactly at the nd threshold $q_0=0$.
The first part of the driving term in Eq.(\ref{eq:1}) turns out to be 
\begin{eqnarray}
< pq\alpha \vert tP \vert \phi > &=& {\delta}_{{\lambda}_0,0} 
\sum_{l_{{\alpha}'}l_0I_0} <pl_{\alpha} \vert t^{\alpha}(-{{3}\over{4m}}q^2)
 \vert {q\over 2}l_{{\alpha}'} > {\varphi}_{{\alpha}_0}(q) \cr
&~& (1m_dI_0m_n\vert JM)~({\lambda}_00{{1}\over{2}}m_n\vert I_0m_n) 
\sqrt{{{\hat{\lambda}_0}\over{4\pi}}}.
\label{eq:a1}
\end{eqnarray}
The quantities with index 0 refer to the initial state. 

The kernel applied on $T$ is given as
\begin{eqnarray}
< pq\alpha \vert tG_0PT &=& \int  {q''}^2dq'' \cr 
&~& \int_{-1}^{+1}dx 
{ {
  \sum_{l_{{\alpha}'}} {{m}\over {qq''} }
{ {
t^{l_{\alpha},l_{{\alpha}'}} (p,{\pi}_1;-{{3}\over {4m}} q^2) 
 }\over {
{\pi}_1^{l_{{\alpha}'} }
} }
\sum_{{\alpha}''} { {G_{\bar {\alpha}{\alpha}''}(q,q'',x)  }
\over {{\pi}_2^{l_{{\alpha}''} }  } } 
<{\pi}_2q''{\alpha}''\vert T
} \over {x_0 - x} 
}
\label{eq:a2}
\end{eqnarray}
with
\begin{eqnarray}
x_0 &\equiv & { {-k_d^2 - q''^2 - q^2 } \over {qq'' } }
\label{eq:a3}
\end{eqnarray}
and $\bar {\alpha}$ contains the same quantum numbers as ${\alpha}$ 
 with the exception of $l_{\alpha}$ replaced by $l_{{\alpha}'}$.

For our notation see~\cite{glo83}. The deuteron binding energy is written as 
$(-k_d^2/m)$. The remaining parts related to $V_4^{(1)}$ 
can be worked out correspondingly and can be found in~\cite{hub}. Evaluating 
the elastic scattering amplitude one needs it at $q=q_0$ 
(see Eq.(\ref{eq:4})). 
Therefore the point $q=q_0=0$ was included. Then Eq.(\ref{eq:a2}) 
simplifies to 
\begin{eqnarray}
< pq=0\alpha \vert tG_0PT &=& 2m {\delta}_{{\lambda}_{\alpha},0} 
\int  {q''}^2dq''   \cr 
&~&
  \sum_{l_{{\alpha}'}} 
{ {
t^{l_{\alpha},l_{{\alpha}'}} (p,q'';0) 
\sum_{{\alpha}''} 2^{l_{{\alpha}''}} 
g_{\bar {\alpha}{\alpha}''}^{0l_{{\alpha}'}0l_{{\alpha}''}0} 
<{1\over 2}q''q''{\alpha}''\vert T
} \over {-k_d^2 - q''^2} 
}.
\label{eq:a4}
\end{eqnarray}

One ends up with the elastic scattering amplitude at threshold:
\begin{eqnarray}
U_{{\lambda}'I',{\lambda}I}^{J^{\pi}} &=&
-{ {2k_d^2} \over {m } }
{\delta}_{{\lambda},0} {\delta}_{{\lambda}',0} 
g_{{\alpha}_{d'}{\alpha}_d}^{00000} \sum_{l,l'} 
{ {{\varphi}_{l'}(p) } \over {p^{l'} } }{\vert}_{p=0}
{ {{\varphi}_{l}(p) } \over {p^{l} } } {\vert}_{p=0} \cr 
&+& {\delta}_{{\lambda}',0} \sum_{l',{\alpha}''} 2^{l''+1} 
g_{{\alpha}_{d'}{\alpha}''}^{0l'0l''0} 
\int {q''}^2dq'' {\varphi}_{l'}(q'') <{1\over 2}q''q''{\alpha}'' \vert T \cr 
&+& \sum_{l'} \int {p'}^2dp' {\varphi}_{l'}(p') 
\lbrace V_4^{(1)}(1+P)\phi + 
V_4^{(1)}(1+P)G_0T {\rbrace}_{{\alpha}_{d'},{\lambda}I}(p',0).
\label{eq:a5}
\end{eqnarray}

The geometrical coefficients $g_{{\alpha}{\alpha}'}^{kl_1l_2{l'}_1{l'}_2}$ 
arise  from the permutation operator P and are given by Eq.(A19) 
in~\cite{glo83}.
 
\acknowledgements
This work was supported by
the Deutsche Forschungsgemeinschaft (J.G.,R.S.),
by the NFS grant No. PHY0070858, and by  the NATO grant PST.CLG.978943. 
R.S. acknowledges financial support of the Foundation for Polish Science.
The authors would like to thank Dr. E. Epelbaum for very constructive
discussions.
The numerical calculations have been performed
on the Cray T90, SV1 and T3E of the NIC in J\"ulich, Germany.

\begin{table}
\caption{Doublet and quartet nd scattering lengths $^2a$ and  $^4a$ 
together with the coherent scattering 
length $b_{nd}$ for different NN potentials and selected combinations 
with different 3NF's. All calculations have been done with $j_{max}=5$. 
The first and second rows give for each potential or potential 
combination the values obtained with np-pp   strong 
potentials with and without EM interactions, respectively (see text 
for explanation). The third and fourth rows for AV18 and CD~Bonn~2000 
potentials and their combinations with 3NF's are the corresponding results 
when the pp strong NN potential is replaced by the nn one (keeping the nn MMI
in case that EMI are included). 
The last column shows our $^3H$ binding 
energies. We also included in the second column 
the cut-off parameter $\Lambda$ for the TM and 
TM99 forces.}  

\begin{tabular}{|c|c|c|c|c|c|} \hline
      {Potential} & ${{\Lambda}/{m_{\pi}}}$ & $^2a$     & $^4a$    & $b_{nd}$ & $E_{^3H}$ \cr
                  &  & (fm)      & (fm)     & (fm)     & (MeV)     \cr
\hline
      CD~Bonn~2000 &     & 0.976   &  6.347  &  6.837  & -7.946 \cr
             &     &  1.011  &  6.324  &  6.833  & -7.989 \cr
             &     &  0.925  &  6.347  &  6.812  & -8.005 \cr
             &     &  0.943  &  6.324  &  6.798  & -8.048 \cr
\hline
  CD~Bonn~2000+TM  & 4.795    &  0.622  &  6.347  &  6.661  & -8.419  \cr
                  & 4.795    &  0.661  &  6.324  &  6.657  & -8.463 \cr
                  & 4.795    &  0.570  &  6.347  &  6.634  & -8.482 \cr
                  & 4.795    &  0.590  &  6.324  &  6.622  & -8.528 \cr
\hline
  CD~Bonn~2000+TM99 & 4.469   &  0.620  &  6.347  &  6.660  & -8.422 \cr
              & 4.469   &  0.658  &  6.324  &  6.656  & -8.466 \cr
              & 4.469   &  0.569  &  6.347  &  6.634  & -8.482 \cr
              & 4.469   &  0.589 &  6.324  &  6.622  & -8.527 \cr
\hline
  CD~Bonn~2000+Urb &   &  0.637  &  6.347  &  6.668  & -8.423 \cr
               &   &  0.674  &  6.324  &  6.664  & -8.467 \cr
               &   &  0.586  &  6.347  &  6.643  & -8.482 \cr
               &   &  0.607 &  6.325  &  6.630  & -8.526 \cr
\hline
  AV18         &   & 1.304 & 6.346  & 7.001   & -7.569 \cr
               &   & 1.319 & 6.326  & 6.988   & -7.606 \cr
               &   &  1.248   &  6.346  &  6.973  & -7.628 \cr
               &   & 1.263    &  6.326  &  6.960  & -7.666 \cr
\hline
  AV18+TM      & 5.215  & 0.614   & 6.346   & 6.656   & -8.478 \cr
               & 5.215  & 0.633   & 6.326   & 6.645   & -8.518 \cr
               & 5.215  &  0.556  &  6.346  &  6.627  & -8.545 \cr
               & 5.215  &  0.575 &  6.326  &  6.616  & -8.584 \cr
\hline
  AV18+TM99    & 4.764  & 0.645   & 6.346   & 6.671   & -8.417 \cr
               & 4.764  & 0.663   & 6.326   & 6.660   & -8.457 \cr
               & 4.764  &  0.587  &  6.346  &  6.643  & -8.482 \cr
               & 4.764  &  0.606   &  6.326  &  6.632  & -8.522 \cr
\hline
  AV18+UrbIX   &   & 0.636   & 6.347   & 6.667   & -8.418 \cr
               &   & 0.654  & 6.326   & 6.656   & -8.458 \cr
               &   &  0.578  &  6.347  &  6.638  & -8.484 \cr
               &   &  0.597  &  6.326  &  6.628  & -8.523 \cr
\hline
  Nijm~I        &   &  1.158   &  6.342  &  6.924  & -7.742 \cr
               &   &  1.190   &  6.321  &  6.919  & -7.782 \cr
\hline
  Nijm~I+TM     & 5.120  &  0.601  &  6.342  &  6.646  & -8.493 \cr
               & 5.120  &  0.638   &  6.321  &  6.643  & -8.535 \cr
\hline
  Nijm~I+TM99   & 4.690  &  0.594  &  6.342  &  6.642  & -8.485 \cr
               & 4.690  &  0.629  &  6.321  &  6.638  & -8.528 \cr
\hline
  Nijm~II       &   &  1.231   &  6.345  &  6.964  & -7.663 \cr
               &   &  1.259   &  6.325  &  6.957  & -7.700 \cr
\hline
  Nijm~II+TM     & 5.072  &  0.598  &  6.345  &  6.647  & -8.500 \cr
               & 5.072  &  0.630  &  6.325  &  6.643  & -8.540 \cr
\hline
  Nijm~II+TM99   & 4.704  &  0.597  &  6.345  &  6.646  & -8.487 \cr
               & 4.704  &  0.627  &  6.325  &  6.642  & -8.527 \cr
\hline
  Nijm~93       &   &  1.196   &  6.343  &  6.944  & -7.672 \cr
               &   &  1.225   &  6.322  &  6.937  & -7.712 \cr
\hline
  Nijm~93+TM    & 5.212  &  0.574  &  6.343  &  6.633  & -8.502 \cr
               & 5.212  &  0.608  &  6.322  &  6.629  & -8.543 \cr
\hline
\end{tabular}
\label{tab1}
\end{table}

\end{document}